\documentclass[notitlepage,
 aip,
 amsmath,amssymb,
reprint,%
]{revtex4-2}

\usepackage[version=3]{mhchem} 
\usepackage{multirow}
\usepackage{threeparttable}
\usepackage{appendix}
\usepackage{braket}
\usepackage{longtable}
\usepackage{color}
\usepackage{graphicx}
\usepackage{bm}
\usepackage[section]{placeins}
\usepackage{soul}
\usepackage{siunitx}
\usepackage[table]{xcolor}
\usepackage{amsmath}

\renewcommand{\thetable}{\arabic{table}}
\usepackage{multirow}
\usepackage{threeparttable}
\usepackage{eucal}

\begin{document}

\preprint{AIP/123-QED}

\title[]{
Fast and Accurate Charge Transfer Excitations \\via Nested Aufbau Suppressed Coupled Cluster\\
}

\author{Harrison Tuckman}
\affiliation{
Department of Chemistry, University of California, Berkeley, California 94720, USA 
}

\author{Eric Neuscamman}
\email{eneuscamman@berkeley.edu}
\affiliation{
Department of Chemistry, University of California, Berkeley, California 94720, USA 
}
\affiliation{Chemical Sciences Division, Lawrence Berkeley National Laboratory, Berkeley, CA, 94720, USA}

\date{\today}

\begin{abstract}
  
Modeling charge transfer well can require treating
post-excitation orbital relaxations and handling
medium to large molecules in realistic environments.
By combining a state-specific correlation treatment
with such orbital relaxations,
Aufbau suppressed coupled cluster
has proven accurate for charge transfer, but, like many coupled
cluster methods, it struggles with large system sizes.
We derive a low-cost Aufbau suppressed second order perturbation
theory and show that, by nesting a small coupled cluster treatment
inside of it, computational cost and scaling are reduced while
accuracy is maintained.
Formal asymptotic costs are dropped from iterative $N^6$
to non-iterative $N^5$ plus iterative $N^3$,
and we test an initial implementation that
can handle about 100 atoms and 800 orbitals
on a single computational node.
Charge transfer excitation energy errors are \textcolor{black}{typically below}
0.1 eV, \textcolor{black}{with an average} 0.25 eV improvement
over $N^6$-cost equation of motion coupled cluster
with singles and doubles.

\end{abstract}

\maketitle


Charge transfer (CT) excitations are at the crux of photochemical efficacy across a wide range of current scientific interests, including photovoltaics,\cite{coropceanu2019charge} photosynthetic complexes,\cite{hustings2022charge} and photocatalysis.\cite{may2024new}
However, traditional linear response approaches such as
time dependent density functional theory (TD-DFT)
\cite{runge1984density,burke2005time,casida2012progress}
and equation of motion coupled cluster (EOM-CCSD)
\cite{rowe1968equations,stanton1993equation,krylov2008equation}
produce significant errors for these excitations
\cite{sobolewski2003ab,dreuw2003long,dreuw2004failure,mester2022charge,kozma2020new,izsak2020single}
due in part to their limited ability
to capture nonlinear orbital relaxation effects.
\cite{subotnik2011communication,herbert2023density}
The challenge of modeling CT is further
exacerbated by system size:
these excitations often occur in medium to large
donor-bridge-acceptor molecules \cite{albinsson2008long}
and can be strongly influenced by solvent and
other environmental effects. \cite{weaver1992dynamical}
To model charge transfer well, a method must capture orbital
relaxations and their effects on electron correlation
while maintaining a low computational cost.
Our recently introduced Aufbau suppressed coupled cluster
(ASCC) approach
\cite{tuckman2024aufbau,tuckman2025improving}
captures both orbital relaxation and its correlation effects,
but, with a cost like the ground state singles and
doubles theory (CCSD), its applicability to CT
was initially somewhat limited.
In this study, we develop a second order perturbative
counterpart to ASCC and, by nesting a small
coupled cluster evaluation inside of it, reduce the
cost and scaling of the approach while maintaining
its accuracy for valence, Rydberg, and CT excitations.

ASCC and its partially linearized cousin (PLASCC)
have a two-fold advantage in handling
post-excitation orbital relaxations that
makes them particularly well positioned to model CT.
\cite{tuckman2024aufbau,tuckman2025improving}
First, they can be constructed on top of a reference
wave function that already captures orbital relaxation
effects at the mean field level.
Second, their full $\hat{T}_1$ singles operators
allow them to fine tune orbital relaxations
in the presence of correlation effects.\cite{thouless1960stability}
The ability to deliver a CCSD-like weak correlation
treatment alongside and informed by these orbital
relaxations has allowed PLASCC to deliver
$\sim$0.1 eV excitation energy errors across a wide variety
of excited states, including Rydberg and CT states.
\cite{tuckman2025improving}
While this accuracy is especially noteworthy for CT,
where EOM-CCSD's errors are \textcolor{black}{$\sim$0.375} eV,
the $N^6$ cost scaling with system size $N$
has so far limited the applicability of ASCC
and PLASCC to relatively small systems.

To lower cost while maintaining accuracy, we propose
to exploit ASCC theory's strong connections to
perturbation theory (PT) to avoid wasting effort
perfecting the treatment of correlation effects that
are simply going to cancel out when evaluating the
excitation energy difference.
Just as ground state CCSD is motivated by PT and intimately
linked to second order M{\o}ller–Plesset (MP2) theory,
\cite{shavitt2009many,nooijen1999combining,bochevarov2005hybrid,bochevarov2006hybrid,shee2024static}
ASCC and especially PLASCC are motivated by PT arguments
\cite{tuckman2025improving}
and, as we will show, share an analogous link
to an excited state second order Aufbau suppressed PT.
Exploiting this link, we will construct a nested ASCC
approach in which the low cost PT
determines which orbitals'
correlation effects are most strongly influenced by the
excitation, and which orbitals' correlations are
sufficiently insensitive to the excitation that
leaving their treatment at the PT level should be safe.
A subsequent ASCC or PLASCC treatment of only the
strongly affected orbitals will then be used to
refine the correlation effects that most strongly
influence the excitation energy.

In order to arrive at a second order excited-state-specific PT, we will begin with the ASCC equations and systematically remove terms contributing at higher perturbative orders, similarly to how MP2 can be derived from ground state CCSD.\cite{shavitt2009many} The ASCC ansatz is exponentially parameterized using both an excitation operator $\hat{T}$ and a one-body deexcitation operator $\hat{S}^{\dagger}$ that act on a closed-shell Aufbau determinant, $\ket{\phi_0}$.\cite{tuckman2024aufbau}
\begin{equation}
    \ket{\Psi _{\text{ASCC}}}=e^{-\hat{S}^{\dagger}}e^{\hat{T}}\ket{\phi_0}
\end{equation}
With this choice of ansatz, remarkably simple initializations of $\hat{T}$ and $\hat{S^{\dagger}}$ allow us to transform $\ket{\phi_0}$ into a qualitatively correct open-shell excited state, at which point $\hat{T}$ can be further optimized to capture weak correlation effects.\cite{tuckman2024aufbau}
We move towards working equations in the same manner as traditional
CC, which leads to similarity transformations of the Hamiltonian.
\begin{align}
    \bar{H}&=e^{-\hat{T}}e^{\hat{S}^{\dagger}}\hat{H}e^{-\hat{S}^\dagger}e^{\hat{T}}
\end{align}
Since $\hat{S}^\dagger$ is a one-body operator, the inner
transformation amounts to a two-electron integral transform.
Working with the outer transformation leads to tensor
contractions similar to those in CCSD, but which are evaluated
using the modified integrals resulting from the inner transform.
As in ground state CC, the ASCC energy is evaluated as the expectation
value with respect to $\ket{\phi_0}$, while projections with excited
determinants $\bra{\phi _\mu}$ yield the amplitude equations that
we solve to optimize $\hat{T}$.
\begin{align}
    E&=\bra{\phi_0}\bar{H}\ket{\phi_0}\label{eqn: energy}\\
    0&=\bra{\phi_\mu}\bar{H}\ket{\phi_0}\label{eqn: residual}
\end{align}

Considering these equations now from a perturbative perspective, the initialization that both $\hat{S}^{\dagger}$ and $\hat{T}$ require to arrive at a qualitatively correct starting point implies that they both contain nontrivial zeroth order pieces. \cite{tuckman2025improving} Note, however, that $\hat{S}^{\dagger}$ remains fixed after initialization, with only $\hat{T}$ varying and
thus containing higher order contributions.
\begin{align}
    \hat{S}^{\dagger}&=\hat{S}^{\dagger (0)}\\
    \hat{T}&=\hat{T}^{(0)}+\hat{T}^{(1)}+\hat{T}^{(2)}+\cdots
\end{align}
After perturbatively partitioning the Hamiltonian
--- see the Supporting Information (SI) ---
its double similarity transform can also
be partitioned by perturbative order.
\begin{align}
    \hat{H}&=\hat{H}^{(0)}+\hat{H}^{(1)}\\
    \bar{H}&=\bar{H}^{(0)}+\bar{H}^{(1)}+\bar{H}^{(2)}+\cdots
\end{align}
Substituting this expansion of $\bar{H}$ into the energy and residual
equations (Eqns.\ \ref{eqn: energy} and \ref{eqn: residual})
allows us to evaluate cluster amplitude and energy contributions
order by order.
By adding together all energetic contributions up to second order, we are thus able to determine the cluster amplitudes necessary to achieve a second order excited-state-specific energy, thus arriving at the second order PT we will need for our nested approach.
\begin{align}
    E_{\text{PT2}}&=E^{(0)}+E^{(1)}+E^{(2)}\label{eqn:ept2}\\
    E^{(n)}&=\bra{\phi_0}\bar{H}^{(n)}\ket{\phi_0}\\
    0&=\bra{\phi_\mu}\bar{H}^{(n)}\ket{\phi_0}\rightarrow \hat{T}_{\mu}^{(n)}   \label{eqn: aspt2res}
\end{align}

While the working equations for this PT are a bit
more complicated than for MP2 due to the presence of
$\hat{T}^{(0)}$, they do closely mirror MP2's computational cost.
Once the two-electron integrals are available in
the molecular orbital (MO) basis, every term in the working
equations scales as $O(N^4)$ or less, as is the
case for MP2.
Even better, the amplitude equations in Eqn.\ \ref{eqn: aspt2res} are linear in the highest order terms --- which at each order are the ones being solved for --- and are small-block block-diagonal within a semicanonical basis, allowing them to be solved in a non-iterative manner (see SI). Therefore, the headline scaling of this PT, in exact asymptotic equivalence to MP2, comes from the non-iterative $O(N^5)$ cost of transforming the two-electron integrals into the MO basis. Compared to the iterative $O(N^6)$ scaling of ASCCSD, this is a substantial improvement in computational efficiency and notably avoids some of the most memory intensive terms found in CCSD and ASCCSD. Even when compared to relatively affordable wave function correlation methods for excited states, such as the iterative $O(N^5)$ scaling CC2\cite{christiansen1995second} and ADC(2)\cite{schirmer1982beyond,trofimov1995efficient,schirmer2004intermediate} methods, the non-iterative $O(N^5)$ scaling of a routine integral transformation represents a significant cost reduction.

\begin{figure*}[t]
    \centering
    \includegraphics[width=\linewidth]{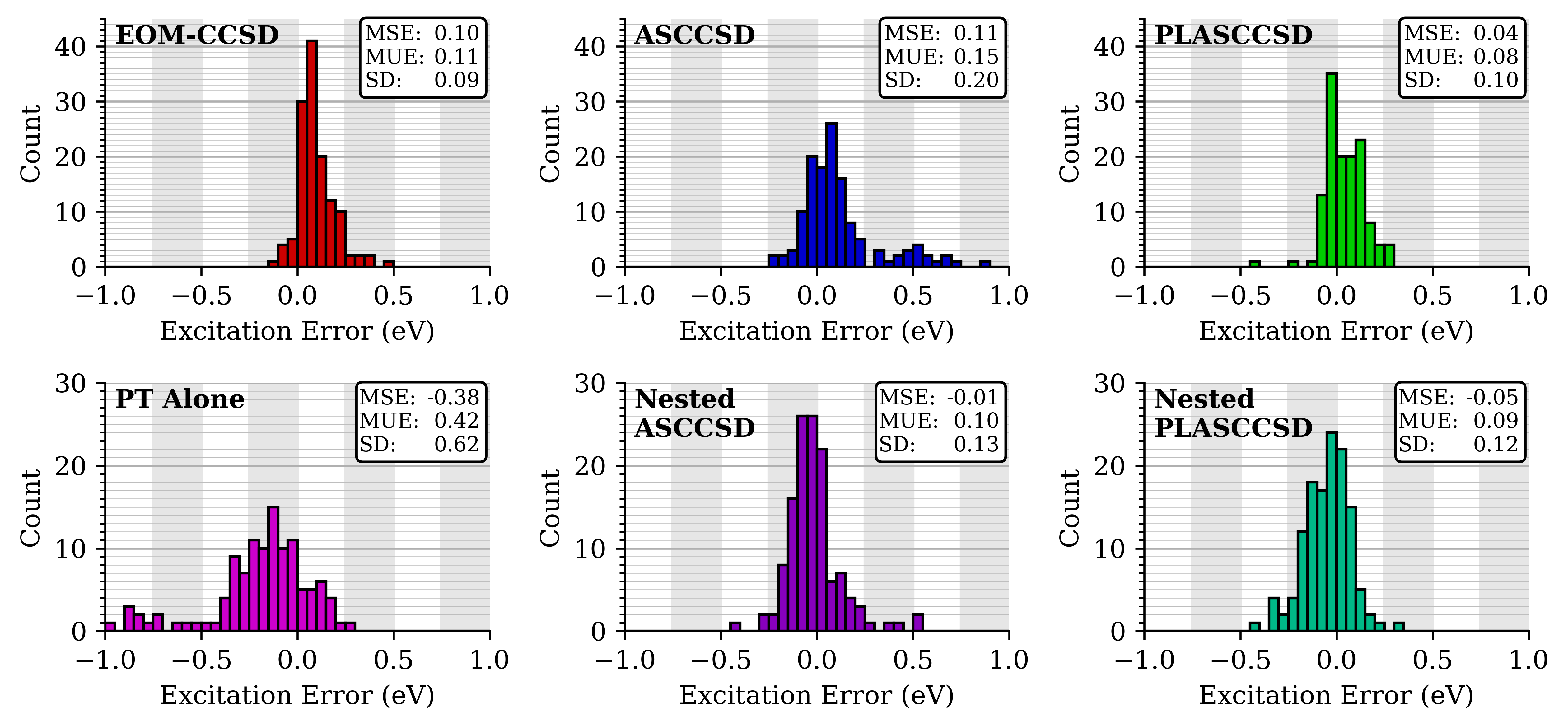}
    \caption{Excitation energy error distributions for various methods on 130 valence and Rydberg single-CSF singlet states from the QUEST benchmark in an aug-cc-pVDZ basis. For the PT alone, 17 outlier states with a $<-1.0$ eV error are not shown. Note the change of scale in the y-axis between the top and bottom rows. The mean signed error (MSE), mean unsigned error (MUE), and standard deviations (SD) are shown in insets. Reference values are of at least EOM-CCSDT quality.
    }
    \label{fig: questhists}
\end{figure*}

\begin{figure}
    \centering
    \includegraphics[width=\linewidth]{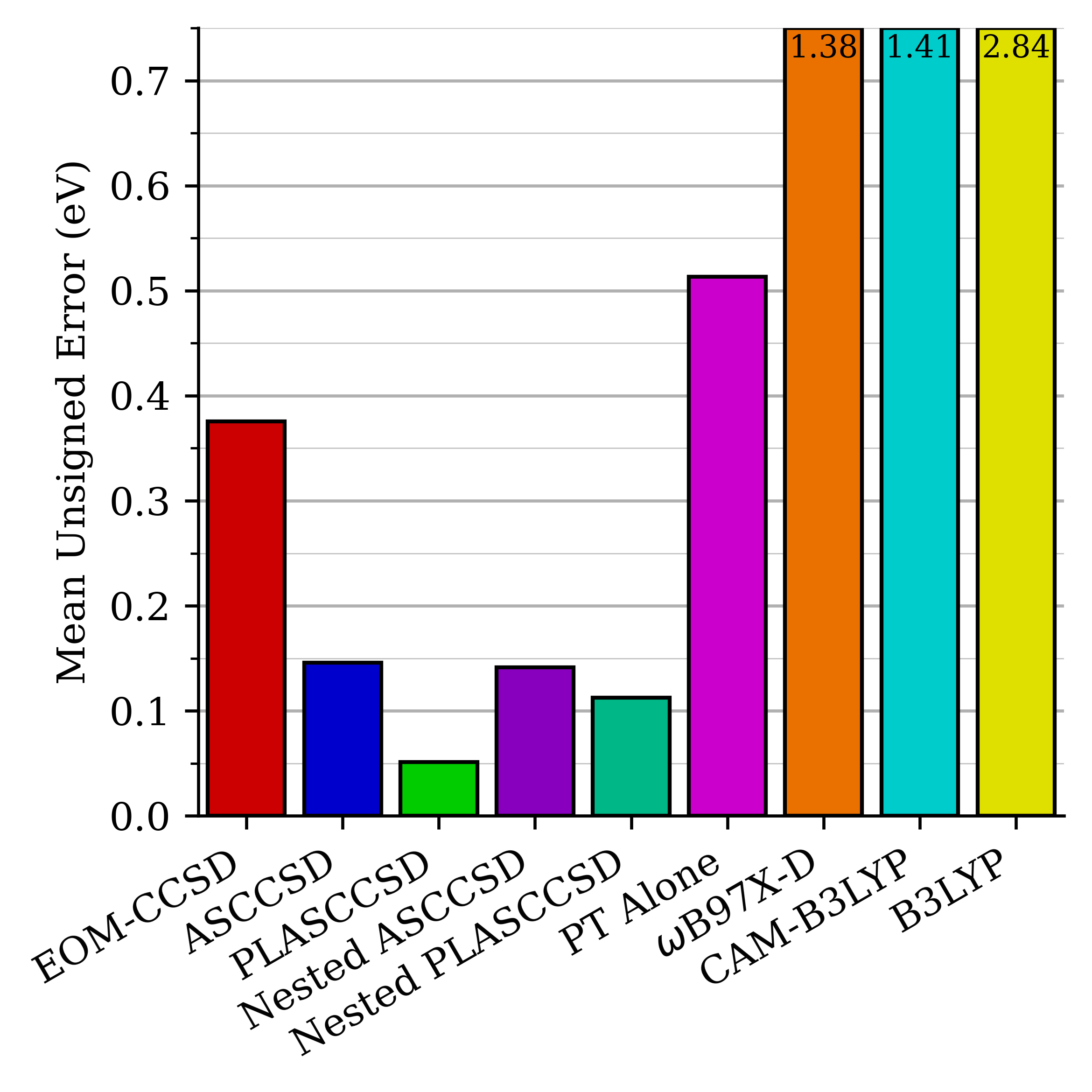}
    \caption{Mean unsigned errors (MUEs) for various methods on \textcolor{black}{16 intermolecular and intramolecular} CT excitations in a cc-pVDZ basis. MUEs extending beyond the plot are printed at the top of each bar. Reference values are EOM-CCSDT except for one LR-CC3 value.
    }
    \label{fig: ctbar}
\end{figure}

Before we go about nesting a small ASCC treatment within this PT, it is worthwhile to evaluate its accuracy as a standalone method. To do so, we test the PT on a collection of 130 singly excited, single CSF, singlet valence and Rydberg states from the QUEST molecular benchmark\cite{loos2018mountaineering,loos2020mountaineering} as well as a set of \textcolor{black}{16} CT states, \textcolor{black}{with intermolecular CT states coming} from the benchmark of Szalay and coworkers\cite{kozma2020new} \textcolor{black}{and intramolecular CT states coming mainly from the long range CT states in the benchmark of Loos and coworkers.\cite{loos2021ct}} The results are summarized in Figures \ref{fig: questhists} and \ref{fig: ctbar}, with additional details available in the SI 
\textcolor{black}{(note that we will develop the nested theories referred to in these figures further down in the main text).}
While computationally efficient, the PT is unsurprisingly less accurate than its parent CC theory.
Though it performs well in many cases, it produces a substantial number of severe errors, some larger than an eV, particularly in the aromatic molecules where PLASCC makes the largest improvements over ASCC.
Indeed, this pattern helps corroborate the original motivation for PLASCC, which was that in these systems the perturbation is large enough that higher order and poorly balanced terms in ASCC become important. \cite{tuckman2025improving}
At the very least, these PT results suggest that the perturbation
in ASCC theory is indeed larger in many of these ring states.

Of particular concern for the present study is that the PT
fares worse in CT states (errors of \textcolor{black}{$\sim$0.5} eV) than it does in
valence and Rydberg states (errors of $\sim$0.4 eV).
This worsening contrasts sharply with PLASCC, which \textcolor{black}{typically}
produces errors of $\sim$0.1 eV in each category.
One culprit is likely the truncation of singles contributions to
low orders in PT theory, which interferes with the parent
theory's ability to provide CC-singles-style corrections
to the initial mean field orbital relaxations.
\cite{tuckman2024aufbau,tuckman2025improving}
Another factor is that more electron pairs
experience large, excitation-induced changes to their
electron correlation in CT states, because CT
excitations make major changes to how close various
electrons are to each other.
It is to be expected that second order PT
would be less effective
at capturing these changes than a CCSD-style approach.
These challenges notwithstanding, the PT does offer
significantly higher accuracy for CT than hybrid
TD-DFT methods, even when long-range-corrected
functionals like $\omega$B97X-D \cite{chai2008long}
and CAM-B3LYP \cite{yanai2004new} are used.

To return accuracy to the CC level while keeping cost in check,
we now turn to nesting a small CC
treatment of the most important correlations within
a larger PT treatment of the rest.
We begin by noting that, even in CT excitations that move
electrons around, the number of individual orbitals that are
directly involved in or strongly affected by the excitation
is not likely to grow with system size, and that these orbitals
tend to be spatially localized (e.g.\ on the donor and acceptor).
This motif suggests that, at least in a large molecule, the vast
majority of the correlation contributions will come from orbitals
only weakly influenced by the excitation.
We might therefore expect such contributions to be about the
same in the ground and excited state and to mostly cancel out
in the excitation energy difference.
If that is indeed the case, it is likely unnecessary to treat all
correlations at the CC level.
Instead, we propose that PT should be sufficient for
most of them while a more targeted CC treatment
handles the few that are strongly affected by the excitation.

We therefore begin by performing full MP2 and Aufbau suppressed PT
calculations on the ground and excited states, respectively.
While these use (semi)canonical bases to ensure
a non-iterative $O(N^5)$ cost, we then proceed
to localize and match orbitals between the
ground and excited state (see SI)
in preparation for the correlation analysis.
\textcolor{black}{(While for simplicity we employ standard Foster-Boys
orbital localization \cite{foster1960canonical}
in this initial investigation, related approaches
\cite{aquilante2009systematic,segarra2015multiconfigurational,kumar2017frozen,mukhopadhyay2023state,manna2025reducedcostequationmotion,mester2017reduced,helmich2011local,hoyvik2017correlated,folkestad2019multilevel,folkestad2024reduced}
suggest that localizations of subsets of the
MP2 or Aufbau suppressed PT natural orbitals
\cite{lowdin1955quantum}
or natural transition orbitals \cite{martin2003natural}
could be even more effective, especially for the virtual orbitals.)}
In this matched local basis, we construct ground and excited
state measures for the correlation contributions related to each orbital.
Specifically, for a given orbital, we add together all of the PT's
amplitude contributions to Eq.\ \ref{eqn:ept2} in which the
amplitude involved has at least one index on that orbital.
We then compare this correlation measure between the ground and the
excited state, and, if they differ by more than a small set threshold,
that orbital gets flagged as needing to be part of the nested CC treatment.
One can imagine other screening approaches, and it's hard to know what
the optimal approach would be, but we will emphasize some strong formal
properties of this approach:
it preserves size consistency, extensivity, and intensivity
and, for a given localized orbital,
is invariant to rotations among the other
localized occupied or virtual orbitals.

Having identified the orbitals whose correlation is
strongly affected by the excitation, we are ready to
perform a small ASCC refinement nestled within the
larger PT treatment.
To do so, we freeze at the PT level all
amplitudes with one or more indices on weakly
affected orbitals and proceed to solve the ASCC
or PLASCC equations for the remaining amplitudes,
using their PT values as an initial guess.
We therefore only need to work with the amplitude residuals
(Eq.\ \ref{eqn: residual}) for which
all residual indices are on strongly affected orbitals.
Once we have solved these residual equations for their
corresponding amplitudes, we take these refined amplitudes
along with those frozen at the PT level and perform one final
whole-system evaluation of the CC energy in the ground
and excited states.

As there are expected to be only $O(1)$ orbitals whose
correlation is strongly affected by the excitation,
this nested approach leads to a dramatic cost reduction
compared to the iterative $O(N^6)$ cost of a
full ASCC evaluation.
We can break the cost of the nested CC into two parts:
the non-iterative part and the iterative part.
The final energy evaluation, as in standard CCSD,
carries a non-iterative $O(N^4)$ cost, and indeed
all non-iterative portions of the nested CC are
$O(N^4)$ or better.
To see why, note that amplitude indices ranging over $O(N)$
values must contract with the Hamiltonian if their amplitude
is to contribute to one of the $O(1)$ residuals we are
evaluating.
As the Hamiltonian remains two-body even after the
$\hat{S}^\dagger$ transformation, it has only four indices to
work with, and so any such term can have no worse than an
$O(N^4)$ cost scaling.

Similar arguments reveal that the iterative portion of the
nested CC need not have a cost worse than $O(N^3)$.
We note that, due to the connectedness of the CC equations,
any amplitude appearing in the residual equations must
contract with the Hamiltonian.
If all four of the Hamiltonian's indices participate in
$O(N)$-range contractions,
then we know that none of the amplitudes are among those
updated in the nested treatment, and so the contribution
in question can be evaluated once and stored,
becoming one of the non-iterative terms.
Thus, there cannot be more than three indices with
$O(N)$ ranges in each term that actually changes during the
iterative solution of the amplitude equations, and so
we conclude that the iterative cost is not worse than $O(N^3)$.
While a careful use of intermediates may be able to drop this iterative scaling
further, in the present study we have focused on ensuring
that our implementation achieves this iterative $O(N^3)$ bound.
Thus, all told, our nested CC implementation involves
non-iterative $O(N^4)$ steps and iterative $O(N^3)$ steps,
while the transformation of the PT amplitudes into the
local basis for the correlation analysis involves
a non-iterative $O(N^5)$ step, much like a standard
two-electron integral transform would.

\begin{figure}
    \centering
    \includegraphics[width=\linewidth, trim=7 16 25 5, clip]{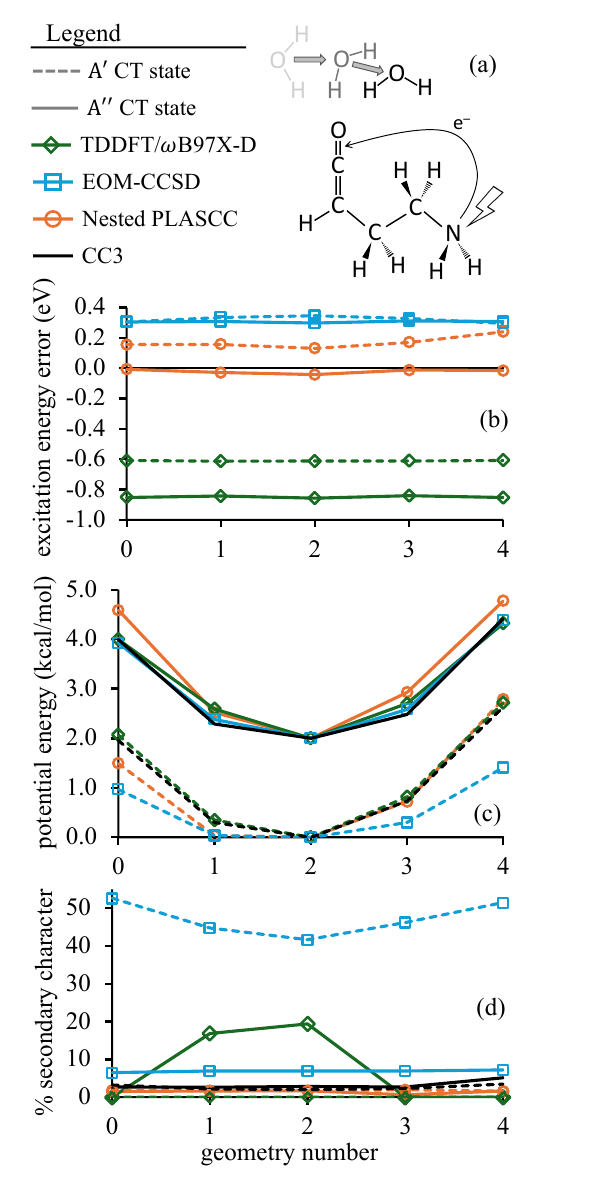}
    \caption{\textcolor{black}{Tests on the ${}^1A'$ (dashed lines) and ${}^1A''$ (solid lines)
    $n\rightarrow\pi^{*}$ CT states as a water is moved near its hydrogen
    bond acceptor.
    Shown at the top (a) is a sketch of the water's positions
    at geometry numbers 0, 2, and 4, from left to right, respectively.
    The three plots show (b) excitation energy errors relative to CC3,
    (c) the excited state potential energy surfaces, whose zeros of
    energy are set to geometry 2 and where the $A''$ surfaces have then
    been shifted up by 2 kcal/mol for clarity, and
    (d) the weight of the largest non-CT component in the state.
    The basis is cc-pVDZ.
    See text and SI for details.}
    }
    \label{fig:flyby}
\end{figure}

\textcolor{black}{While our nesting approach
and double similarity transform contain parallels
to multilevel CC (MLCC) theory \cite{myhre2014multi}
and similarity transformed EOM (STEOM),
\cite{nooijen1997similarity}
respectively, it is important to appreciate the differences.
MLCC, like EOM-CCSD, employs the equation-of-motion formalism for excited states,
\cite{myhre2014multi,hoyvik2017correlated,folkestad2019multilevel}
and so it will inherit EOM-CCSD's difficulties with post-excitation
orbital relaxations.
ASCC and PLASCC, in contrast, provide a more robust orbital relaxation
treatment, as discussed above.
This difference explains their accuracy advantage over EOM-CCSD
in CT states, an advantage that we will see is
maintained even when nested inside the new PT.
In STEOM, electron affinity and ionization potential calculations are used
to construct the second similarity transform,
\cite{nooijen1997similarity}
which for long-range CT should
build in the appropriate orbital relaxations. However, it is less obvious that
the same will be true for local excitations, where
no region of the molecule loses or gains an electron.
In ASCC, the second similarity transform is constructed from
an orbital-relaxed, neutral excited state reference,
and any further fine tuning of the relaxation can be done
state-specifically via the exponentiated singles operator.
}

We now turn to the question of whether the nested
approach maintains the accuracy of its parent CC theory.
Figure \ref{fig: questhists} reveals that, for valence and Rydberg states,
the nested approach is far more accurate than the PT alone and, at least for PLASCC,
has an accuracy commensurate with the full CC treatment.
Interestingly, nested ASCC is \textit{more} accurate than full ASCC, which
is likely driven by the same effect that makes PLASCC more accurate than ASCC.
We observed previously \cite{tuckman2025improving} that certain nonlinear
terms in ASCC are less well balanced than in the ground state due to
the perturbative incompleteness of their energy contributions.
PLASCC counteracts this effect through a partial linearization that explicitly
deletes the offending terms, whereas nested ASCC appears to at least partially
counteract it by freezing many of the amplitudes at a specific order in
PT theory, where such imbalances do not exist because all terms contributing
at a given order are included.
As we saw when testing PLASCC, \cite{tuckman2025improving}
this effect can be observed clearly in nested ASCC's tendency to
reduce symmetry breaking relative to full ASCC.
In addition to good accuracy, these tests also reveal that,
even in the modestly sized molecules in this benchmark, the cost reduction from
nesting is significant.
While the number of occupied orbitals whose correlation contributions are
frozen at the PT level in these small molecules is roughly what one would
expect from the frozen core approximation,
the number of virtual orbitals with PT-frozen correlation
is quite substantial, typically 40\% or more of the virtual set.

Turning now to the CT results in Figure \ref{fig: ctbar},
we see a similarly encouraging story unfold.
Nested ASCC and PLASCC are much more accurate than the
PT theory on its own.
PLASCC and nested PLASCC are again more accurate than ASCC,
although in this case nested ASCC is not a \textcolor{black}{significant} improvement
over full ASCC.
\textcolor{black}{For all but two states, nested PLASCC closely maintains PLASCC's accuracy, ultimately resulting in a}
0.25 eV improvement over EOM-CCSD, which is now especially significant in
light of the non-iterative $O(N^5)$ vs
iterative $O(N^6)$ computational costs.
Nested PLASCC far outstrips the TD-DFT accuracy of
the density functionals we test here, and is also
a notable improvement over the much wider
range of TD-DFT methods tested by
Mester and K\'allay. \cite{mester2022charge}
\textcolor{black}{With these initial surveys
confirming nested PLASCC's accuracy
in valence, Rydberg, and CT states,
we will now explore an example at the
intersection of CT and hydrogen bonding.}

\textcolor{black}{In the system shown in Figure \ref{fig:flyby}(a),
a water molecule is moved past its low-energy hydrogen
bonding arrangement with the acceptor moiety's oxygen atom.
This displacement test, the geometries for which are
taken from another recent PT study,
\cite{clune2025emlc}
allows us to examine the behavior of 
different methods' excitation energies,
potential energy surfaces (PESs), and state characters as the local
environment of a CT acceptor is modified by a 
non-innocent solvent effect.
We look in particular at the ${}^1A'$ and ${}^1A''$
CT states that promote from the nitrogen lone pair
to the in-plane and out-of-plane $\pi ^*$ orbitals of the CCO group,
respectively.
The remaining panels of Figure \ref{fig:flyby} reveal that,
among the three methods tested against CC3,
nested PLASCC is the only method that simultaneously offers
reasonably accurate excitation energies,
potential energy surfaces within 1 kcal/mol of CC3,
and accurate state characters for both states at all five geometries.
As usual, TD-DFT severely underestimates the CT excitation 
energies, although it does prove quite accurate for
the excited state PESs.
The latter observation is somewhat surprising in light
of the fact that it erroneously mixes a large portion
of $\pi\rightarrow$Rydberg character into the $A''$
state at some geometries, seemingly a result of its too-low
excitation energy causing two states to appear nearly degenerate
when in fact they are not.
Similarly, EOM-CCSD incorrectly mixes a large Rydberg component
into the $A'$ state at all geometries.
In that case, however, the
erroneous state character is sufficient to significantly 
alter the PES, which underestimates the energy needed to
move the water from geometry 2 to geometry 4 by more than 1 kcal/mol.
This underestimation makes sense, as, with half the state not
putting negative charge on the acceptor, it becomes too easy to pull the
partially positive hydrogen atom away from it.
Although this is only one test, it does
suggest that PES explorations may be
a useful application of nested PLASCC
once its analytic gradients have
been implemented.
For now, we will turn our attention to a more pressing
concern, which is whether the theoretically expected
cost scaling of the approach is seen in practice.
}

\begin{figure}
    \centering
    \includegraphics[width=\linewidth]{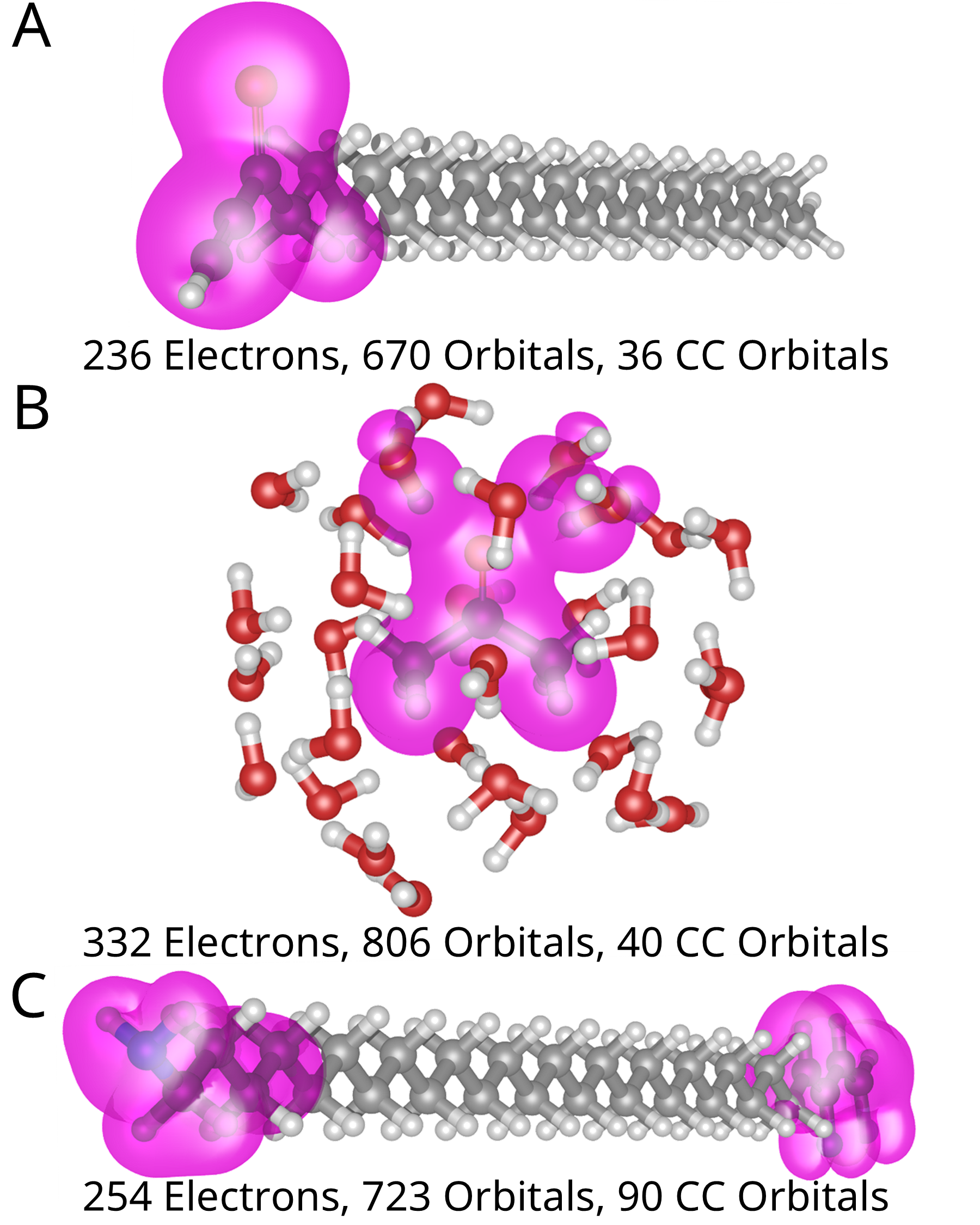}
    \caption{A gas phase, simple valence excitation (A), a simple valence excitation in a solvated environment (B), and a gas phase, donor-bridge-acceptor CT excitation (C) computed with the nested ASCC approach. Pink surfaces are 95\% isosurfaces for the electronic density of the occupied orbitals receiving a CC correlation treatment, as automatically determined by the
    PT's correlation analysis.}
    \label{fig: surface}
\end{figure}

\begin{figure}
    \centering
    \includegraphics[width=\linewidth]{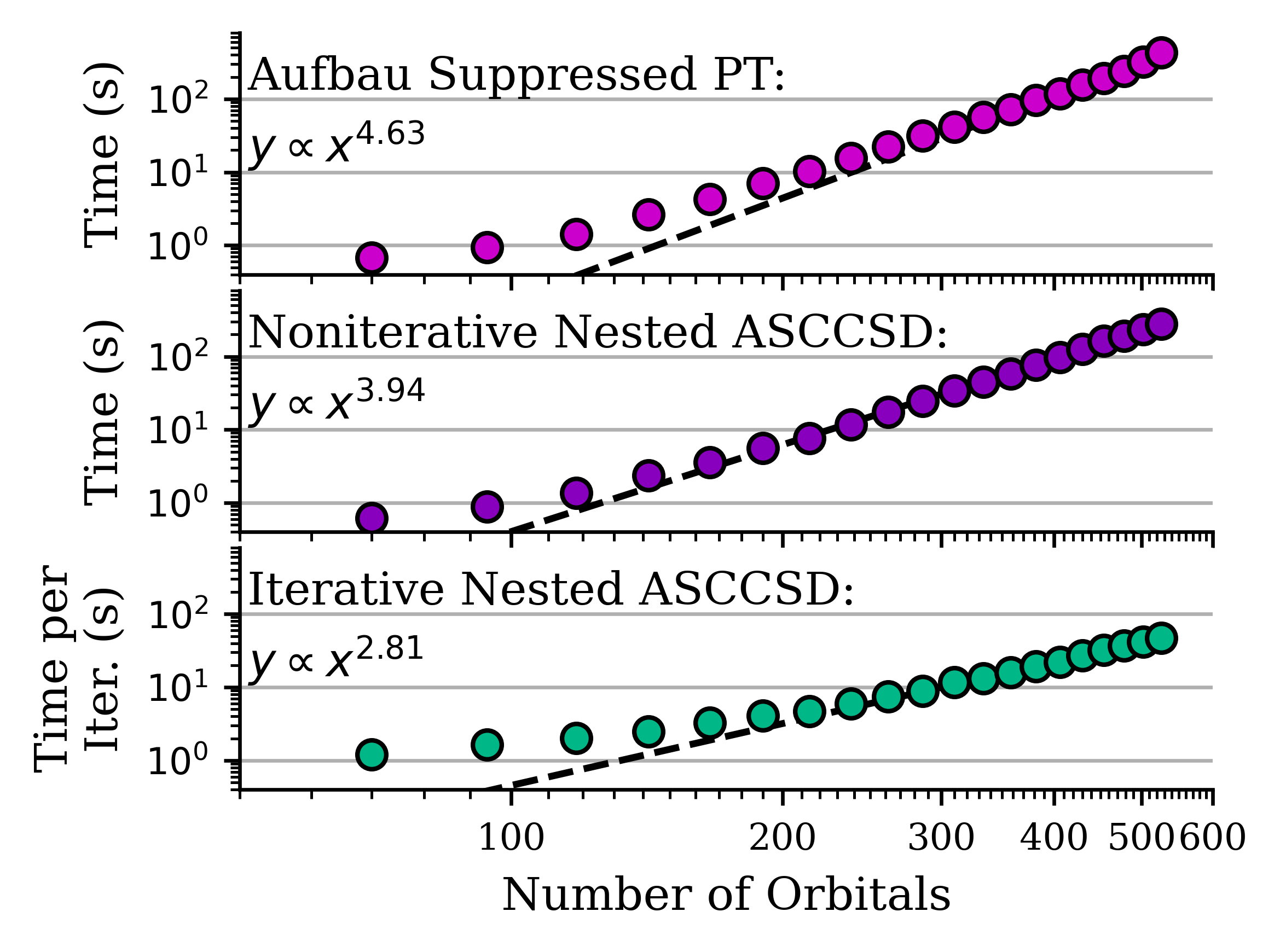}
    \caption{Wall time measured for Aufbau Suppressed PT (top), the noniterative step in nested ASCCSD (middle), and the iterative step in nested ASCCSD (bottom) on one Intel Xeon Gold 5218 processor containing 32 cores for the $n$ to $\pi ^*$ excitation in thiopropynal with a varying length alkane chain. Dashed lines show best fit curves of the form $y\propto x^{a}$ fit to the last 10 data points.}
    \label{fig: timing}
\end{figure}

We begin with the simple gas phase valence \textcolor{black}{$n\rightarrow\pi ^*$} excitation in \textcolor{black}{a thiopropynal-based molecule with an extended alkane chain}
(Figure \ref{fig: surface}A) to demonstrate a few key points.
First, this excitation is an example where the nested CC refinement
makes a significant, $\sim$0.15 eV difference compared to the PT alone.
This difference is likely driven by the importance of
post-excitation orbital relaxations on the relatively polarizable sulfur atom
following the delocalization of the lone pair electron into the $\pi^*$ system.
Second, despite this delocalization, the excitation remains localized
compared to the extent of the molecule, which should leave most orbitals'
correlation contributions mostly unaffected.
Finally, its computational cost grows with system size
as we would expect it to, as seen in Figure \ref{fig: timing}.

For an explicit solvation example, we examine in Figure \ref{fig: surface}B
the $n\rightarrow\pi^*$ transition in solvated acetone.
The explicit waters play a non-innocent role, as the hydrogen bonding that
initially stabilizes the ground state is disrupted when one of the
hydrogen-bond-accepting lone pair electrons is transferred
into the $\pi^*$ orbital by the excitation.
Crucially, despite the strong solute-solvent interactions and the
three-dimensional nature of this example, the orbitals
flagged for CC refinement remain few in number and are
tightly localized in space.
We find it especially noteworthy that, despite no explicit input to this
effect from the user, the PT's correlation analysis selected CC orbitals
in a physically intuitive manner: the water orbitals it picks out
are the ones that had their strong hydrogen bonding interactions
perturbed by the excitation.

Finally, we examine the donor-bridge-acceptor CT example of
Figure \ref{fig: surface}C, which results from adding an
alkane bridge to one of our earlier CT tests.
\textcolor{black}{In this state, a lone pair electron on the nitrogen donor is transferred to the $\pi^*$ system on the opposite end of the molecule.}
In contrast to the last two examples in which the excitation was
localized in only one region of space, here the PT's correlation
analysis flags two separate regions for CC treatment:
one on the donor and one on the acceptor, again following what
physical intuition would suggest.
While this donor/acceptor effect leads to a larger number of
orbitals in the CC calculation, this number stops changing
with system size once the bridge reaches just three carbons
in length.
Thus, regardless of whether the excitation itself is a local valence
excitation or a non-local CT, we see that the nesting approach
successfully demarcates a physically intuitive set of orbitals
for careful CC treatment while dramatically reducing the
computational cost compared to a full-system CC evaluation.

In conclusion, we have developed an approach for replicating
the high accuracy of excited-state-specific Aufbau suppressed coupled
cluster at low cost by nesting a small coupled cluster evaluation
within a larger perturbative treatment.
This new perturbation theory displays strong parallels to MP2,
including the headline $N^5$ cost scaling from its bottleneck
integral transformation and its strong connections to coupled
cluster theory.
We find it especially interesting that a simple analysis of
this perturbation theory's correlation contributions is able to
automatically demarcate physically intuitive orbital subsets for
which a coupled-cluster-level treatment is most essential when
predicting excitation energies.
While the perturbation theory is not particularly accurate on its own,
the overall nesting approach rivals the parent PLASCC method in
accuracy in valence, Rydberg, and CT states.
This accuracy is especially promising given the approach's
headline cost reduction from iterative $N^6$ to non-iterative $N^5$
and the dramatically larger system sizes that can be reached by
an initial implementation.

While this nesting approach is a significant step forward, it is only one
step in a larger journey towards high-accuracy, low-cost models of
charge transfer excitations.
Given the strong parallels between the new perturbation theory and MP2,
as well as those between ASCC and CCSD,
it seems likely that local correlation techniques will further reduce
both the scaling and cost of this approach.
Accuracy can likely be improved as well, as the same perturbative framework
used here should also be effective for deriving higher-order non-iterative
perturbative corrections to the ASCC energy, as has been so richly
explored in the ground state.
These continuing advances promise to bring ever larger molecules and
more complicated chemical environments within reach of high-accuracy
excited state modeling.
\section{Computational Methods}
\textcolor{black}{
Our EOM-CCSD calculations and Foster-Boys orbital localizations\cite{foster1960canonical} were performed with PySCF,\cite{sun2015libcint,sun2018pyscf,sun2020recent}
our DFT and EOM-CCSDT calculations were performed with Q-Chem,\cite{epifanovsky2021software} and our LR-CC3 calculations were performed with PSI4.\cite{smith2020psi4} Though the literature and reference calculations were performed utilizing the frozen core approximation, our EOM-CCSD, PT, and ASCC calculations were performed without this approximation. Nevertheless, this is expected to make only a small difference to the excitation energies ($\sim$0.02 eV)\cite{loos2018mountaineering,loos2020mountaineering} and thus the high accuracy frozen-core values still make excellent references. The geometries for the large molecule valence and charge transfer excitations were determined via
MP2/cc-pVDZ geometry optimizations in Q-Chem, while the solvated acetone geometry was determined using the solvator method in ORCA.\cite{neese2025software,bannwarth2019gfn2}
Explicit geometries can be found in the SI.}

\textcolor{black}{
For the excited state PT, ASCC, and PLASCC calculations, the excited state reference was determined via the excited state mean field (ESMF) method,\cite{shea2018communication,shea2020generalized,hardikar2020self} which can be summarized as a CIS wavefunction with full orbital relaxations. Any CSF in the ESMF reference with a singular value greater than 0.2 was considered part of the excited state reference.
Block-sparse \cite{rubensson2005systematic} Cholesky decompositions
\cite{koch2003reduced}
with a threshold of $10^{-7}$ Hartree
were used to evaluate MO basis two electron integrals,
with Libint\cite{Libint2} supplying the underlying
atomic orbital integrals.
For the ASCC based approaches, convergence was considered achieved when the maximum residual was no greater than $10^{-5}$ and the energy changed by less than $10^{-7}$ Hartree in an iteration. For all other methods, default convergence criteria were used. On rare occasions, convergence difficulties were encountered for ASCC based approaches. Details regarding these states may be found in the SI.
}

\textcolor{black}{
For the nested ASCC calculations, three orbital energy thresholds for
establishing the small orbital set for the CC part were tested.
The benchmark data in the main text are with a 0.005 eV threshold,
with additional accuracy data for 0.01 eV and 0.0025 eV thresholds
available in the SI.
For the test in Figure \ref{fig:flyby}, the number of CC occupied and virtual orbitals was fixed with the tighter 0.0025 eV threshold at the equilibrium geometry and enforced across all other geometries on the surface in order to promote smoothness, and the excited state absolute energy was determined by adding the PT2 absolute energy of the ground state to the nested PLASCC excitation energy.
Finally, the scaling analysis was performed using the 0.01 eV threshold.
}

\section{Acknowledgements}

This work was supported by the National Science Foundation,
Award Number 2320936.
Calculations were performed 
using the Savio computational cluster resource provided by the Berkeley Research Computing program at the University of California, Berkeley and the Lawrencium computational cluster resource provided by the IT Division at the Lawrence Berkeley National Laboratory.
H.T. acknowledges that this 
material is based upon work supported by the National Science Foundation 
Graduate Research Fellowship Program under Grant No.\ 
DGE 2146752. Any opinions, findings, and conclusions or recommendations 
expressed in this material are those of the authors and do not necessarily 
reflect the views of the National Science Foundation.

\section{References}
\bibliographystyle{achemso}
\bibliography{main}

\clearpage
\onecolumngrid

\section{Supporting Information}
\renewcommand{\thesection}{S\arabic{section}}
\renewcommand{\theequation}{S\arabic{equation}}
\renewcommand{\thefigure}{S\arabic{figure}}
\renewcommand{\thetable}{S\arabic{table}}
\setcounter{section}{0}
\setcounter{figure}{0}
\setcounter{equation}{0}
\setcounter{table}{0}
\section{Perturbatively Partitioning the Hamiltonian}

\begin{figure}
    \centering
    \includegraphics[width=\linewidth]{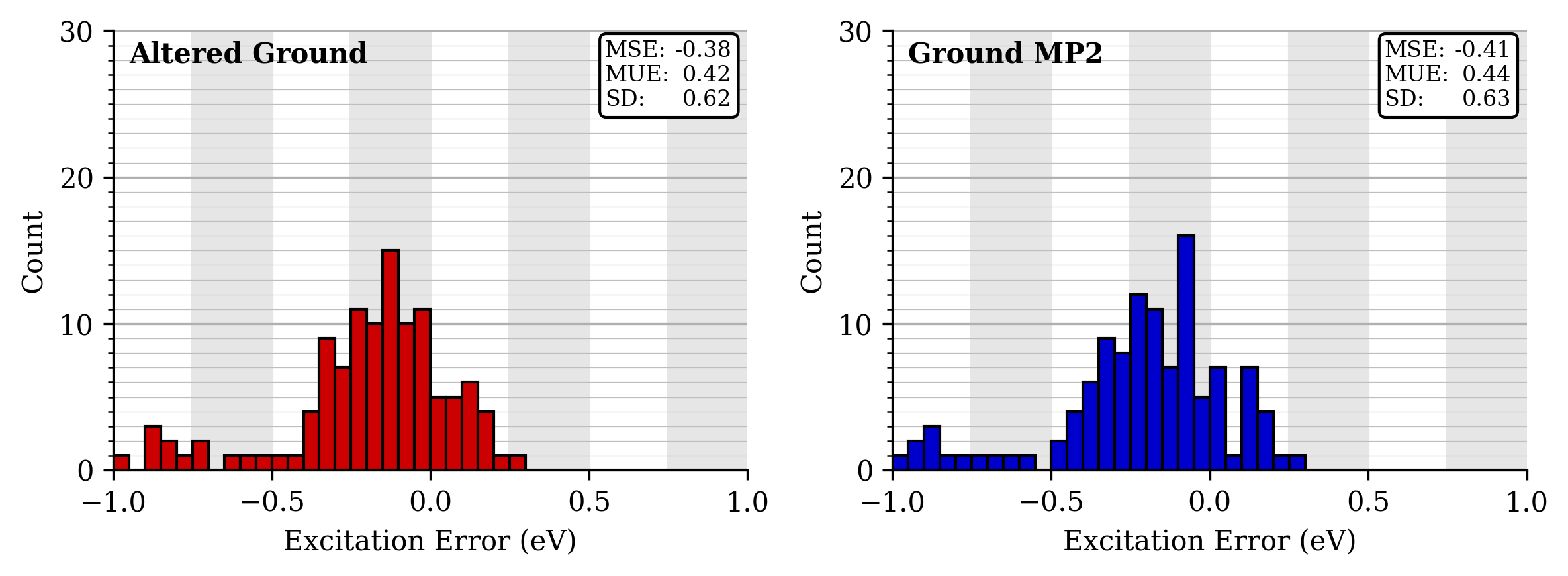}
    \caption{Error distributions for the Aufbau Suppressed PT's excitation energy predictions on the QUEST set when using either the altered zeroth order Hamiltonian (left) or the traditional MP2 choice (right) in the ground state calculation. Mean signed errors (MSEs), mean unsigned errors (MUEs), and standard deviations (SD) are shown for both methods.}
    \label{fig:mp2comp}
\end{figure}

In order to utilize perturbation theory, a choice must be made as to how to partition the Hamiltonian. In the ground state MP theory with a general single determinant (not necessarily Hartree Fock), the zeroth order Hamiltonian is chosen to be the occupied-occupied and virtual-virtual blocks of the one-electron Fock operator while the first order Hamiltonian is everything else.
\begin{align}
    \hat{H}&=H^{(0)}+H^{(1)}\\
    \hat{H}^{(0)}&=\sum_{ij}f_{ij}\hat{i}^\dagger \hat{j}+\sum_{ab}f_{ab}\hat{a}^\dagger \hat{b}\\
    \hat{H}^{(1)}&=\hat{H}-\hat{H}^{(0)}
\end{align}
In the Aufbau Suppressed PT used in this study, an additional further subdivision must be made in order to delineate the importance of the primary hole and particle orbitals being excited from and into during a particular excitation.
Thus, we will now take the indices $ij\cdots$ and $ab\cdots$ to represent \textit{nonprimary} occupied and virtual orbitals, respectively.
$h$ and $p$ will refer to hole and particle orbitals,
respectively.
\begin{align}
    \hat{H}&=H^{(0)}+H^{(1)}\\
    \hat{H}^{(0)}&=\sum _{h_1h_2}f_{h_1 h_2}\hat{h}_1^\dagger\hat{h}_2+\sum_{ij}f_{ij}\hat{i}^\dagger \hat{j}\notag\\
    &\quad+\sum _{p_1p_2}f_{p_1 p_2}\hat{p}_1^\dagger\hat{p}_2+\sum_{ab}f_{ab}\hat{a}^\dagger \hat{b} \label{eqn:pt-H0-detail}\\
    \hat{H}^{(1)}&=\hat{H}-\hat{H}^{(0)}
\end{align}
Though slightly more complicated, this form of $\hat{H}^{(0)}$ maintains a block diagonal structure and ensures that the only zeroth order cluster amplitudes are those mentioned in the main text: the subset of amplitudes initialized to nonzero values during the Aufbau-suppression-style construction of the qualitatively correct starting point. When this starting point is a single CSF, as was true for all states in this study, 
the hole and particle sums in Eq.\ \ref{eqn:pt-H0-detail} each contain just one term, because in these cases there are just one hole orbital and one particle orbital. In order to maintain equivalence between the perturbation theories of the ground and excited states, this partitioning scheme was also utilized for the ground state, which causes the ground state energy to differ slightly from the usual MP2 energy. However, this choice does slightly improve the PT's accuracy, as can be seen in Figure \ref{fig:mp2comp}.

\clearpage
\section{Solution to the PT Equations}

In the canonical orbital basis, the diagonality of the Fock operator allows the MP2 linear equation for the amplitudes
to be solved in a non-iterative manner, as it guarantees
that each linear equation is fully decoupled from the others. In our PT, while many of the linear equations are also decoupled in the same way when expressed in a semicanonical basis, the presence of the off-diagonal, deexciting zeroth order Hamiltonian elements in conjunction with zeroth order cluster amplitudes causes some of the linear equations to be coupled. As an example of these coupled equations, the highest order perturbative amplitude terms are shown for the all-primary singles and doubles residual equations in the semicanonical basis for the single-CSF case.

\begin{align}
    \bar{H}^{(0)}&=e^{\hat{S}^\dagger}\hat{H}^{(0)}e^{-\hat{S}^\dagger}\notag\\
    &=\sum _{h_1h_2}\bar{f}_{h_1 h_2}\hat{h}_1^\dagger\hat{h}_2+\sum_{ij}\bar{f}_{ij}\hat{i}^\dagger \hat{j}\notag\\
    &\quad+\sum _{p_1p_2}\bar{f}_{p_1 p_2}\hat{p}_1^\dagger\hat{p}_2+\sum_{ab}\bar{f}_{ab}\hat{a}^\dagger \hat{b}\notag\\
    &\quad+\sum _{hp}\bar{f}_{hp}\hat{h}^{\dagger}\hat{p}
\end{align}

\begin{align}
    {R_h^p}^{(n)}&=\left(\bar{f}^{(0)}_{pp}-\bar{f}^{(0)}_{hh}-2\bar{f}_{hp}^{(0)}{t_h^p}^{(0)}\right){t_h^p}^{(n)}+\bar{f}_{\bar{h}\bar{p}}^{(0)}{t_{h\bar{h}}^{p\bar{p}}}^{(n)}+\cdots\\
    {R_{h\bar{h}}^{p\bar{p}}}^{(n)}&=\left(-2\bar{f}_{hp}^{(0)}{t_h^p}^{(n)}-2\bar{f}^{(0)}_{\bar{h}\bar{p}}{t_{\bar{h}}^{\bar{p}}}^{(n)}\right){t_{h\bar{h}}^{p\bar{p}}}^{(0)}
    \\&+\left(\bar{f}_{pp}^{(0)}+\bar{f}_{\bar{p}\bar{p}}^{(0)}-\bar{f}_{hh}^{(0)}-\bar{f}_{\bar{h}\bar{h}}^{(0)}-2\bar{f}_{hp}^{(0)}{t_h^p}^{(0)}-2\bar{f}_{\bar{h}\bar{p}}^{(0)}{t_{\bar{h}}^{\bar{p}}}^{(0)}\right){t_{h\bar{h}}^{p\bar{p}}}^{(n)}+\cdots
\end{align}

If the number of equations in a coupled block scaled with system size, a non-iterative solution would become prohibitively expensive, and iterative methods would be required. However, because the off-diagonal zeroth order Hamiltonian element deexcites only in the small primary orbital space, the blocks of coupled equations are small and independent of system size. In the example shown above, because $\bar{H}^{(0)}$ contains exactly one unique off-diagonal element, only two linear equations are coupled. In the single-CSF case, the largest blocks contain just 6 coupled linear equations.
Thus, solving the PT's amplitude equations amounts to
explicitly inverting $O(N^4)$ small matrices, none of which
have dimensions that grow with system size.
Although a bit more complicated than the fully diagonal
MP2 equations, the cost of solving our PT's equations is also
non-iterative $O(N^4)$.
Thus, as for MP2, the bottleneck in large systems
is actually the $O(N^5)$ cost of transforming the
two-electron integrals into the molecular orbital basis.

\clearpage
\section{Orbital Transformations for\\Nested Coupled Cluster}

Upon completion of the perturbation theory calculations, the orbitals are in their respective semicanonical bases for both the ground and excited states, which may differ due to post-excitation orbital relaxations.
However, in order to exploit locality when determining which correlations are and are not strongly affected by the excitation, we prefer to transform from the semicanonical basis into a local orbital basis.
To do so, the excited state's non-hole occupied and
non-particle virtual orbitals are separately localized
using the Foster-Boys method.
Note that we leave the hole and particle orbitals unchanged
in order to maintain the compactness of the determinant
representation of the excited state.
At this point, we construct matching ground state orbitals
by projecting each of the excited state's (hole + localized)
occupied orbitals into the span of the ground occupieds,
after which a L\"owdin orthonormalization provides an
orthonormal set of ground state occupied orbitals whose shapes
are closely matched to the excited state occupieds.
We repeat this process for the (particle + localized)
excited state virtual orbitals to produce a similarly matched
ground state virtual basis.

Now that we have localized ground and excited state orbitals
whose shapes and therefore physical interpretations are
similar, we perform the correlation
analysis described in the main text that separates
the molecular orbitals into two sets: those that will
have their correlation frozen at the PT level and
those that will have it refined via CC.
Note that, to perform this analysis, we need to first
rotate the PT amplitudes into the local basis,
which incurs a non-iterative $O(N^5)$ cost similar
to that of a standard two electron integral transformation.
Finally, to improve the approximate Jacobian for the
iterative solution of the nested CC treatment,
the set of non-hole and non-particle orbitals
flagged for CC treatment are returned to a
semicanonical basis, again leaving the hole and particle
untouched to maintain the compact form of the excited state.
The PT2 amplitudes are rotated into this new basis as well,
this time at just an $O(N^4)$ cost thanks to savings
related to the $O(1)$ range of indices for orbitals
receiving the CC refinement.

\clearpage
\section{QUEST and Charge Transfer Benchmark}

The full table of results for the QUEST and charge transfer benchmarks may be found in 
the included .xlsx file.
On rare occasion, an ASCC-based method would experience convergence difficulties, particularly when the ESMF wave function had significant deficiencies as compared  to EOM-CCSD. If neither of the two solutions would converge for any of the ASCC-based methods, they were entirely removed for all methods. States where only one of the solutions failed to converge or only a subset of methods failed to converge are highlighted in red. For the states where only one of two solutions would converge, the final reported energy is not the arithmetic mean of the two solutions but just the energy of the only converged solution. Finally, a handful of states would stall near but not quite at our tight convergence criterion. These states, highlighted in blue, are still included due to their energy changing below the level of energetic precision reported in this work.

\begin{figure}
    \centering
    \includegraphics[width=\linewidth]{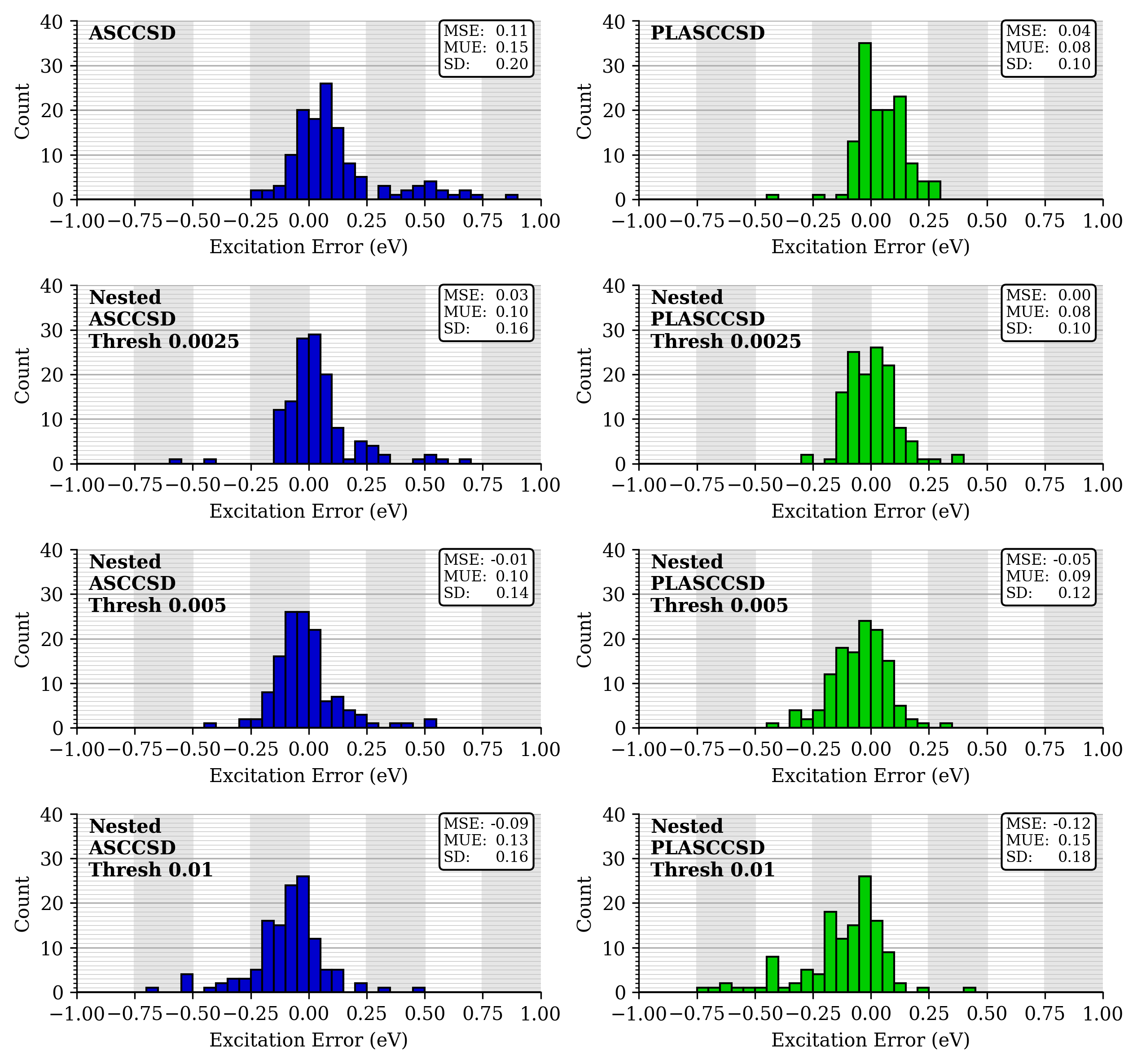}
    \caption{Excitation energy error distributions for ASCCSD (left) and PLASCCSD (right) with differing nesting orbital energy thresholds in eV on 130 valence and Rydberg single-CSF singlet states from the QUEST benchmark in an aug-cc-pVDZ basis. The mean signed error (MSE), mean unsigned error (MUE), and standard deviations (SD) are shown at the top of each plot. Reference values are of at least of EOM-CCSDT quality.}
    \label{fig:thresh}
\end{figure}

Three different energetic thresholds for the nesting procedure were tested in this study: 0.0025 eV, 0.005 eV, and 0.01 eV.
The accuracy for each of the three tested thresholds for the QUEST benchmark are summarized in Figure \ref{fig:thresh}.
Ultimately, we decided to report the results of the 0.005 eV threshold in the main text due to its balance of computational efficiency and accuracy\textcolor{black}{, though we note that the optimal threshold will likely depend on the specifics of the molecular system, excited state character, and basis set.  We look forward to clarifying
the specifics of this dependence in future work.}

\clearpage
\section{State Characterization for Potential Energy Surfaces}

\textcolor{black}{
In an effort to characterize how accurately each method reproduces the state character of the two CT states in the hydrogen bonding test, we have constructed a rough measure for state characterization.
The idea is to quantify the largest component of the state that is not CT, and to compare that between methods, but since we are working with significantly different methods implemented in different codes that print different amounts of state information, a somewhat crude approach is necessary.
For the $A'$ state, we define the state's ``secondary character'' (i.e.\ the biggest component that is not CT) as (a) the weight of the largest printed single excitation other than the nitrogen-lone-pair-to-$\pi ^*$ excitation divided by (b) the sum of the weights of all printed single excitations.
For the $A''$ state, most methods recovered the $\pi^*$ orbital from a linear combination of multiple virtuals, meaning that the CT character of the state was constructed using a superposition of multiple single excitations.
We therefore define this state's secondary character as (a) the net weight of all printed single excitations whose holes were not the nitrogen lone pair divided by (b) the net weight of all printed singles. 
Although the resulting comparison is necessarily somewhat approximate, it does make clear that, in two cases, one for TD-DFT and one for EOM-CCSD, a state was erroneously predicted to contain a substantial amount of non-CT character.
As seen in the figure in the main text, both CC3 and
nested PLASCC predict that both states have only very
small amounts of secondary character.
}

\clearpage
\section{Geometries}

\textcolor{black}{
Information regarding geometries for all molecules in the valence and Rydberg QUEST benchmarks as well as the intermolecular CT states and one intramolecular CT state may be found in the Supporting Information from our previous work.
\cite{tuckman2025improving}
Geometries for the remaining intramolecular CT states may be found in the Supporting Information of the work of Loos and coworkers.
\cite{loos2021ct}
Geometries for the hydrogen bonding test may be found in the Supporting Information of the work of Clune and coworkers.
\cite{clune2025emlc}
The remaining geometries for the large molecule tests may be found below in Angstroms.
}

\vspace{3mm}

\noindent \bf{Thiopropynal alkane chain}
\begin{verbatim}
H     17.22333472     -2.23915362     -0.03663289
C     16.36633614     -1.58788420     -0.03067667
C     15.40552368     -0.81011800     -0.02633299
C     14.24962842      0.05208865     -0.01618250
S     14.42365780      1.68894198     -0.01589282
C     12.91978288     -0.68129515     -0.00697016
H     12.92848499     -1.35261850      0.87545130
H     12.91948877     -1.35965812     -0.88389531
C     11.66900178      0.19160285     -0.00424797
H     11.68786240      0.85658715      0.87801085
H     11.68400899      0.85596281     -0.88708278
C     10.38560742     -0.64441549     -0.00112851
H     10.37917591     -1.30863308      0.88500802
H     10.37493287     -1.30843929     -0.88736459
C      9.11714123      0.21356443      0.00190063
H      9.12851765      0.87741843      0.88793853
H      9.12478282      0.87802384     -0.88373285
C      7.82807710     -0.61299393      0.00426540
H      7.82108794     -1.27678016      0.89068617
H      7.81854533     -1.27784227     -0.88133820
C      6.55946725      0.24485440      0.00552339
H      6.56936720      0.90981034      0.89097232
H      6.56686993      0.90870733     -0.88078266
C      5.26949163     -0.58047861      0.00785620
H      5.26181733     -1.24374203      0.89466527
H      5.26007678     -1.24603874     -0.87721558
C      4.00110337      0.27769660      0.00799701
H      4.01047827      0.94294096      0.89328147
H      4.00920427      0.94135501     -0.87849006
C      2.71074150     -0.54706737      0.00967009
H      2.70234819     -1.21027482      0.89651853
H      2.70162926     -1.21274847     -0.87531687
C      1.44253653      0.31138820      0.00899141
H      1.45155074      0.97677192      0.89418982
H      1.45125791      0.97493918     -0.87758547
C      0.15194430     -0.51302222      0.01006396
H      0.14294281     -1.17616866      0.89695349
H      0.14314317     -1.17879214     -0.87486074
C     -1.11611231      0.34565446      0.00865385
H     -1.10742491      1.01115966      0.89377061
H     -1.10685346      1.00909373     -0.87800746
C     -2.40685576     -0.47852214      0.00920465
H     -2.41635790     -1.14161178      0.89613182
H     -2.41536065     -1.14436111     -0.87567130
C     -3.67480369      0.38031421      0.00716275
H     -3.66641268      1.04592432      0.89220630
H     -3.66509941      1.04364994     -0.87957398
C     -4.96564168     -0.44371458      0.00726944
H     -4.97554446     -1.10675676      0.89422771
H     -4.97387240     -1.10960787     -0.87756811
C     -6.23352221      0.41521780      0.00469499
H     -6.22539802      1.08091953      0.88967337
H     -6.22346537      1.07846062     -0.88210844
C     -7.52440638     -0.40873460      0.00443568
H     -7.53461588     -1.07173877      0.89141850
H     -7.53239042     -1.07466925     -0.88037250
C     -8.79224801      0.45024505      0.00142652
H     -8.78434603      1.11602839      0.88634664
H     -8.78191546      1.11340746     -0.88543487
C    -10.08314236     -0.37367391      0.00087930
H    -10.09355970     -1.03665338      0.88787788
H    -10.09090111     -1.03964250     -0.88390493
C    -11.35095771      0.48530969     -0.00246838
H    -11.34321376      1.15117049      0.88239762
H    -11.34040538      1.14840853     -0.88937736
C    -12.64178754     -0.33864485     -0.00322562
H    -12.65220455     -1.00162487      0.88377478
H    -12.64923185     -1.00465307     -0.88798482
C    -13.90965475      0.52018363     -0.00681707
H    -13.90191061      1.18612668      0.87800134
H    -13.89884300      1.18324380     -0.89376604
C    -15.20013217     -0.30380880     -0.00770684
H    -15.21071702     -0.96721258      0.87924697
H    -15.20754489     -0.97026532     -0.89240151
C    -16.46892214      0.55354390     -0.01145302
H    -16.45910676      1.21879769      0.87251008
H    -16.45590538      1.21579763     -0.89762463
C    -17.75044497     -0.28328228     -0.01234868
H    -17.79682534     -0.93188381      0.87947822
H    -17.79355516     -0.93497942     -0.90208078
H    -18.65132508      0.35250411     -0.01510695
\end{verbatim}

\noindent \bf{Solvated acetone}
\begin{verbatim}
C    -0.00002055      0.00000337      0.09952080
C     0.00000339      1.28065824     -0.69730806
C     0.00001356     -1.28066977     -0.69722662
O     0.00000782     -0.00000034      1.31378426
H    -0.00000092      2.13624078     -0.02743871
H    -0.00000131     -2.13623626     -0.02743654
H     0.87877589      1.31456782     -1.34406044
H    -0.87877280      1.31458555     -1.34406760
H     0.87874031     -1.31456662     -1.34411046
H    -0.87874537     -1.31458273     -1.34410233
O     3.75699034     -2.39641269     -0.67104103
H     3.29242149     -3.02350603     -0.09471475
H     3.35701473     -2.50844957     -1.55546127
O    -1.46613260     -2.16747491      2.42602027
H    -2.13989653     -1.62494532      2.87737626
H    -0.91436065     -1.49616908      1.97932927
O    -2.06307331     -0.04994330     -2.51987743
H    -2.68550831     -0.77157770     -2.75792350
H    -2.21400256      0.70375335     -3.12322156
O    -0.39749759     -4.58945221      1.57960600
H    -0.83865144     -3.82937019      1.99606011
H    -0.82238260     -4.68696654      0.71065809
O    -0.87506662      1.67241888      3.21074846
H    -0.53527183      1.21738144      2.40374454
H    -0.20922266      1.40137853      3.86557873
O     2.78396403     -0.00098899     -0.14091159
H     2.59235067      0.03438425     -1.10403582
H     3.27811538     -0.83996407     -0.05482333
O     2.50955567      2.03144873      1.67581953
H     3.34166806      2.42923564      1.41080855
H     2.39578488      1.26749806      1.07915836
O     2.14865106     -3.86370678      1.08778891
H     2.19209564     -3.11255424      1.71032446
H     1.27775815     -4.26871810      1.29995701
O    -2.54662755      2.73528028      1.21455956
H    -2.24313392      2.67711470      2.12779301
H    -2.47889014      1.83196632      0.84913988
O    -0.10663065      1.29332862     -5.15697583
H     0.53229011      1.90074111     -4.73844880
H    -0.97449004      1.59052173     -4.84742509
O    -2.35311310      2.25249902     -3.78660554
H    -1.74582387      2.95327039     -3.47146674
H    -3.10410099      2.33966628     -3.18214335
O    -3.14523032      0.32155588     -0.08352288
H    -4.00083304      0.70728830     -0.28848599
H    -2.68628391      0.26287884     -0.95950042
O     2.32984344     -2.69097528     -3.04302740
H     1.57041244     -2.18253034     -3.37076484
H     1.94287277     -3.47480428     -2.59897174
O    -3.09505244     -2.41492926      0.21412972
H    -3.17719438     -1.44710700      0.09444861
H    -2.62643153     -2.50223382      1.06552146
O     1.73807430      4.22122825     -1.38805105
H     2.45929751      3.70586631     -0.98339746
H     1.01865970      4.25490694     -0.73909786
O    -1.48641651     -4.29292014     -0.96878531
H    -2.09784333     -3.64910229     -0.55990415
H    -1.37825985     -3.97860092     -1.88787867
O     1.27574640     -4.63985768     -1.51178453
H     1.73789465     -4.50067543     -0.67203326
H     0.33232945     -4.61673361     -1.27946455
O     0.19449960     -0.89241286     -3.75972991
H     0.14555589     -0.27352558     -4.52049571
H    -0.53492918     -0.56626047     -3.17938972
O    -0.94718565     -3.35391858     -3.48674418
H    -0.55117449     -2.47228705     -3.65946608
H    -0.25420709     -3.98816961     -3.68070520
O     2.44247635      0.27044678     -2.82870740
H     3.13548562     -0.26170003     -3.22683992
H     1.59600545     -0.14758969     -3.12626202
O     1.69809111      2.76967698     -3.70299374
H     2.05249070      1.91041745     -3.40144345
H     1.71093661      3.33815996     -2.90783828
O     1.52925386     -1.80282831      2.77575581
H     1.06571218     -1.22970636      2.13413613
H     0.83003164     -2.27091485      3.24663809
O    -3.01403612     -0.03290087      2.80499922
H    -3.15234958      0.12089592      1.85814871
H    -2.33607449      0.61303398      3.08300678
O     1.55022139      0.72662949      3.90763753
H     1.97403659      1.25818048      3.21103400
H     1.71297858     -0.19647479      3.64967223
O     0.63568174      3.86416901      2.45934525
H     0.01150064      3.22830992      2.86005479
H     1.36819745      3.29966007      2.14806820
O    -3.41714064     -2.27523543     -2.66102162
H    -3.47619917     -2.50866811     -1.72327927
H    -2.70954915     -2.81839502     -3.04042042
O     3.75142779      2.39874438     -0.95363915
H     3.56904846      1.51818981     -0.56632444
H     3.69998361      2.24857566     -1.90255218
O    -0.65647136      4.45095076      0.20501945
H    -0.13703428      4.48419830      1.03618362
H    -1.43184228      3.90844229      0.45792820
O    -0.99609885      4.26182040     -2.55548624
H    -0.85762000      4.42376263     -1.60626952
H    -0.12550104      4.36906667     -2.95032900
O    -3.44394726      2.97425463     -1.37841687
H    -2.80691111      3.63599260     -1.67337555
H    -3.41210445      3.02334141     -0.41450195
\end{verbatim}

\noindent \bf{Donor-bridge-acceptor charge transfer}
\begin{verbatim}
C     13.88569453      0.42596960     -0.00000000
H     13.80214752      1.51838629     -0.00000000
C     15.13477199     -0.09201976     -0.00000000
F     15.28098945     -1.43855420     -0.00000000
C     16.36931012      0.68230307     -0.00000000
H     16.26044750      1.77013029     -0.00000000
C     17.63484516      0.21619576     -0.00000000
H     18.50918975      0.87353576     -0.00000000
F     17.95987257     -1.08424358     -0.00000000
C     12.63419683     -0.41586840     -0.00000000
H     12.63471357     -1.08569488      0.88128524
H     12.63471357     -1.08569489     -0.88128524
C     11.35731032      0.42964186     -0.00000000
H     11.35973473      1.09272777      0.88653442
H     11.35973473      1.09272777     -0.88653442
C     10.08042779     -0.41347143      0.00000000
H     10.08183874     -1.07770683      0.88589687
H     10.08183874     -1.07770683     -0.88589687
C      8.80131760      0.42827754      0.00000000
H      8.80127553      1.09260870      0.88597105
H      8.80127553      1.09260869     -0.88597105
C      7.52232239     -0.41379318      0.00000000
H      7.52274423     -1.07825958      0.88584186
H      7.52274423     -1.07825958     -0.88584186
C      6.24268284      0.42732199      0.00000000
H      6.24235899      1.09181366      0.88589363
H      6.24235899      1.09181366     -0.88589363
C      4.96384164     -0.41503441      0.00000000
H      4.96441521     -1.07959099      0.88582386
H      4.96441521     -1.07959099     -0.88582386
C      3.68368466      0.42532130      0.00000000
H      3.68301436      1.08985586      0.88586681
H      3.68301436      1.08985586     -0.88586681
C      2.40520350     -0.41759308      0.00000000
H      2.40605070     -1.08216961      0.88582303
H      2.40605070     -1.08216961     -0.88582303
C      1.12464043      0.42215011      0.00000000
H      1.12368841      1.08670291      0.88585444
H      1.12368841      1.08670291     -0.88585444
C     -0.15353019     -0.42123955      0.00000000
H     -0.15245217     -1.08582128      0.88582424
H     -0.15245217     -1.08582128     -0.88582424
C     -1.43440024      0.41803865      0.00000000
H     -1.43557329      1.08259892      0.88584898
H     -1.43557329      1.08259892     -0.88584898
C     -2.71231486     -0.42573995      0.00000000
H     -2.71105111     -1.09032114      0.88582591
H     -2.71105111     -1.09032114     -0.88582591
C     -3.99341772      0.41318321      0.00000000
H     -3.99476235      1.07774623      0.88584666
H     -3.99476235      1.07774623     -0.88584666
C     -5.27112845     -0.43090252      0.00000000
H     -5.26972556     -1.09547952      0.88582774
H     -5.26972556     -1.09547952     -0.88582774
C     -6.55239575      0.40776698      0.00000000
H     -6.55386460      1.07232922      0.88584688
H     -6.55386460      1.07232922     -0.88584688
C     -7.82995153     -0.43654686      0.00000000
H     -7.82845954     -1.10111554      0.88582985
H     -7.82845954     -1.10111554     -0.88582985
C     -9.11131138      0.40197251      0.00000000
H     -9.11285721      1.06652855      0.88585179
H     -9.11285721      1.06652856     -0.88585178
C    -10.38873073     -0.44252235     -0.00000000
H    -10.38716968     -1.10708305      0.88582618
H    -10.38716968     -1.10708305     -0.88582618
C    -11.67016855      0.39584089      0.00000000
H    -11.67181364      1.06037715      0.88586700
H    -11.67181364      1.06037715     -0.88586700
C    -12.94728798     -0.44905351     -0.00000000
H    -12.94555550     -1.11359752      0.88583384
H    -12.94555550     -1.11359751     -0.88583384
C    -14.22878741      0.38900672     -0.00000000
H    -14.23083524      1.05343330      0.88589108
H    -14.23083524      1.05343330     -0.88589108
C    -15.50541254     -0.45669273     -0.00000000
H    -15.50193862     -1.12156464      0.88565040
H    -15.50193862     -1.12156464     -0.88565040
C    -16.78778519      0.37930436     -0.00000000
H    -16.79500245      1.04342061      0.88698173
H    -16.79500245      1.04342062     -0.88698173
C    -18.06717389     -0.46559851     -0.00000000
H    -18.06532569     -1.12984475      0.88383629
H    -18.06532569     -1.12984475     -0.88383629
N    -19.33029431      0.28206183     -0.00000000
H    -19.32122636      0.91307888      0.80713882
H    -19.32122636      0.91307888     -0.80713882
\end{verbatim}

\renewcommand{\thesection}{\Roman{section}}
\setcounter{section}{6}

\end{document}